\documentclass[conference]{IEEEtran}
\IEEEoverridecommandlockouts
\usepackage{cite}
\usepackage{amsmath,amssymb,amsfonts}
\usepackage{algorithmic}
\usepackage{graphicx}
\usepackage{subfigure}
\usepackage{threeparttable}
\usepackage{setspace}
\usepackage{makecell}
\usepackage[table,xcdraw]{xcolor}
\usepackage{tabularx}
\usepackage{textcomp}
\usepackage{xcolor}
\usepackage{stfloats}
\usepackage{amsfonts}
\usepackage{amssymb}
\usepackage{amsthm}
\usepackage{cite}
\usepackage{float}
\usepackage{color}
\usepackage{stfloats,fancyhdr}
\usepackage{amsmath,bm}
\usepackage{algorithm}
\usepackage{multirow}
\usepackage{changepage}
\usepackage[normalem]{ulem}
\usepackage{balance}

\def\BibTeX{{\rm B\kern-.05em{\sc i\kern-.025em b}\kern-.08em
    T\kern-.1667em\lower.7ex\hbox{E}\kern-.125emX}}

\usepackage{mathtools}
\mathtoolsset{showonlyrefs=true}    
\usepackage{comment}

\begin{document}

\title{\huge Deep Reinforcement Learning for Radio Resource Allocation in NOMA-based Remote State Estimation \vspace{-0.5cm} \\
\thanks{*Wanchun Liu is the corresponding author.}
}

\author{
	\IEEEauthorblockN{	Gaoyang Pang, Wanchun Liu*, Yonghui Li, and Branka Vucetic}
\IEEEauthorblockA{School of Electrical and Information Engineering, The University of Sydney, Australia \\
Emails: \{gaoyang.pang, wanchun.liu, yonghui.li, branka.vucetic\}@sydney.edu.au.}
\vspace{-0.5cm}
}

\maketitle

\begin{abstract}
Remote state estimation, where many sensors send their measurements of distributed dynamic plants to a remote estimator over shared wireless resources, is essential for mission-critical applications of Industry 4.0. Most of the existing works on remote state estimation assumed orthogonal multiple access and the proposed dynamic radio resource allocation algorithms can only work for very small-scale settings. In this work, we consider a remote estimation system with non-orthogonal multiple access. We formulate a novel dynamic resource allocation problem for achieving the minimum overall long-term average estimation mean-square error. Both the estimation quality state and the channel quality state are taken into account for decision making at each time.
The problem has a large hybrid discrete and continuous action space for joint channel assignment and power allocation. We propose a novel action-space compression method and develop an advanced deep reinforcement learning algorithm to solve the problem. Numerical results show that our algorithm solves the resource allocation problem effectively, presents much better scalability than the literature, and provides significant performance gain compared to some benchmarks.
\end{abstract}


\section{Introduction}\label{sec:intro}
Wireless networked control systems (WNCSs), consisting of spatially distributed plants, sensors, machines, actuators and controllers, play an essential role in the era of Industry 4.0 \cite{Park2018WNCS}.  In particular, remote state estimators for monitoring dynamic plant status in a real-time manner are critical in WNCSs to enable high-quality closed-loop control. In Industry 4.0, massive wireless sensors are deployed  for remote state estimation of spatially distributed plants. Thus, it is essential to manage the limited wireless radio resources for remote state estimation of many distributed plants.

Existing works on wireless resource allocation mainly focused on data-oriented communications, and the design targets are transmission throughput, latency, and reliability~\cite{Uysal2021Semantic}. Advanced data-driven machine learning approaches, such as supervised learning and reinforcement learning, have been adopted when resource allocation problems cannot be solved effectively by conventional model-based methods~\cite{Zappone2019DRLoverview}. Different from data-oriented communications, the resource allocation design in a remote estimation system should be task-oriented as the  goal is to minimize the long-term average remote estimation mean-square error (MSE) of the dynamic plant states~\cite{Uysal2021Semantic}. 

Significant efforts have been devoted to the wireless resource allocation of remote estimation systems with orthogonal multiple access (OMA) (see \cite{liu2021deep,Leong2020OMA} and the references therein), where each frequency channel (i.e., a subcarrier) can only be assigned to a single sensor to avoid inter-user interference completely. Non-orthogonal multiple access (NOMA)~\cite{Islam2017NOMA5G} allows simultaneous transmission of multiple sensors' packets  at the same frequency channel and offers higher transmission capacity than OMA. Each sensor packet can be detected by processing received superimposed signals.

However, existing works on dynamic resource allocation in NOMA-based remote estimation only considered the simple multi-sensor-single-channel setting, and focused on either power allocation~\cite{Li2019NOMA2,Forootani2022NOMA,Pezzutto2021NOMA} or channel assignment problems~\cite{Pezzutto2022NOMA}, rather than joint design ones.
Moreover, the developed policy optimization methods for resource allocation can only handle very small-scale systems, e.g., a five-sensor-single-channel setting~\cite{Forootani2022NOMA}, due to the curse of dimensionality in policy optimization.
We aim to address these limitations of NOMA-based remote estimation. The novel contributions of our work are summarized below. 

\textbf{1) New system model.} 
We investigate a NOMA-based remote estimation system with multiple sensors and frequency channels. It requires a joint design of channel assignment and power control of each sensor for achieving the optimal overall remote estimation performance. Such a system has not been investigated before.

\textbf{2) Novel problem formulation.}
We formulate the dynamic resource allocation problem into a Markov decision process (MDP) problem. It takes into account both estimation quality states and channel quality states for decision making. 
To the best of our knowledge, such a problem has not been considered in the literature of remote state estimation. The action space of the formulated MDP is a multi-dimensional hybrid one consisting of discrete channel assignment and continuous power allocation, while classical MDP solutions only work for discrete action spaces. The new (enlarged) state space and the hybrid action space make the decision-making problem difficult to solve, especially in large systems. 

\textbf{3) Advanced dynamic resource allocation algorithm with large state and action spaces.}
To handle the optimal resource allocation problem, we develop an advanced data-driven deep reinforcement learning (DRL) algorithm that generates low-dimensional continuous virtual actions, and propose a novel action mapping scheme to map virtual actions into real hybrid actions in resource allocation.
Extensive simulation results illustrate that the proposed DRL algorithm can effectively solve the dynamic resource allocation problem with a much larger scale than the literature, and provides significant performance gain compared with some benchmark policies, especially when the system scale is large.

\section{System Model of Remote State Estimation} \label{sec:sys}
\begin{figure}[t]
	\centering\includegraphics[width=2.8in]{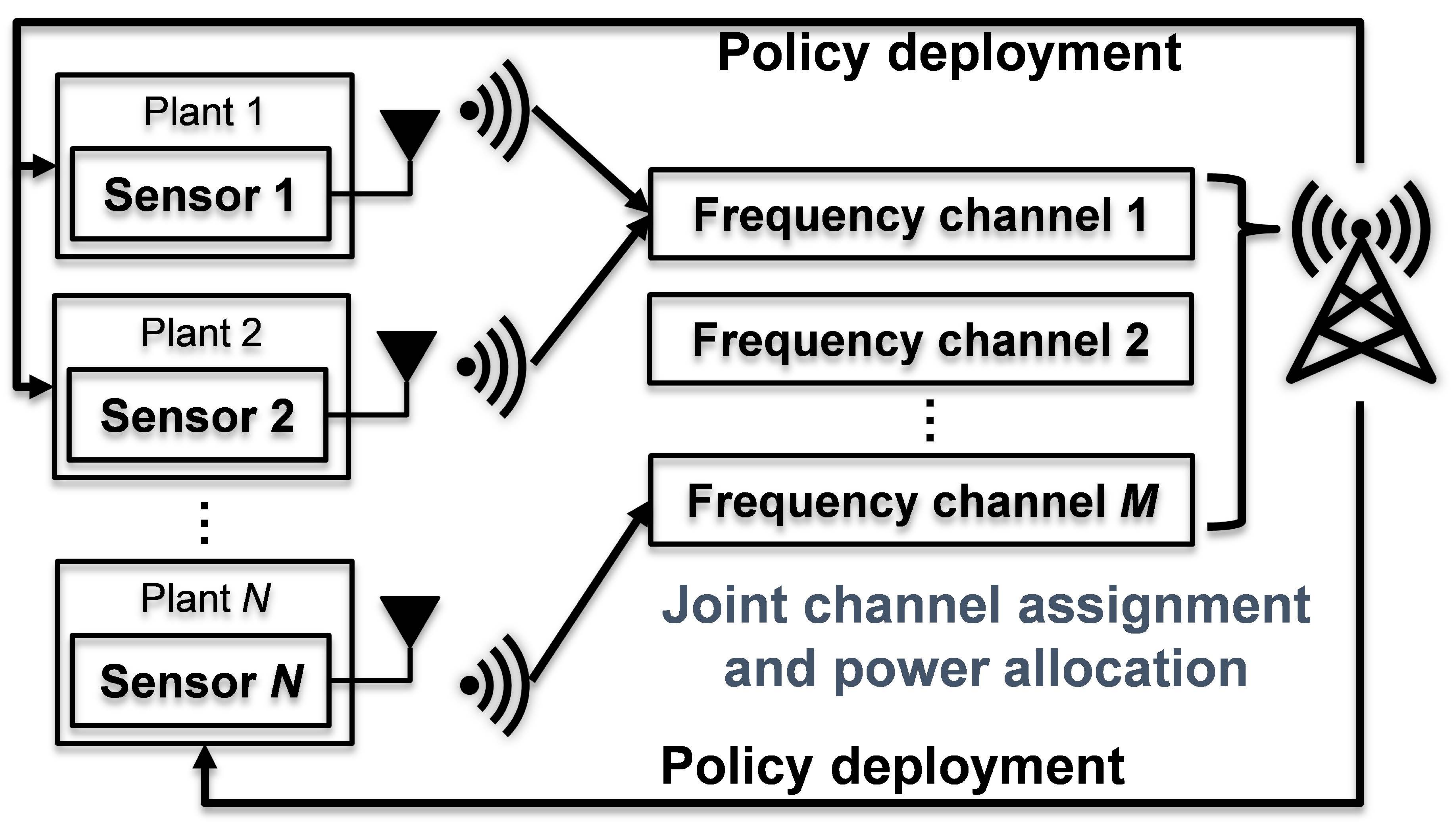}
	\vspace{-0.3cm}
	\caption{A NOMA-based $N$-sensor-$M$-channel remote estimation system.}
	\label{fig:system_model}
	\vspace{-1.5em} 
\end{figure}

The studied remote estimation system consists of $N$ dynamic plants, each monitored by a sensor, and a remote estimator, as shown in Fig.~\ref{fig:system_model}. The $N$ sensors send the measurement data to the remote estimator over $M$ frequency channels (i.e., subcarriers), where $M<N$. Each wireless device is equipped with a single antenna. The remote estimator applies a dynamic resource allocation policy for channel assignment and power control of each sensor.

\subsection{Local State Estimation}
We consider discrete-time linear time-invariant (LTI) systems, where the model of plant $n$ is given as \cite{Leong2020OMA,Liu2021FSMC} 
\begin{equation}\label{LTI}
    \begin{aligned}
    \mathbf{x}_{n}(t+1) &=\mathbf{A}_{n} \mathbf{x}_{n}(t)+\mathbf{w}_{n}(t)\\
    \mathbf{y}_{n}(t) &=\mathbf{C}_{n} \mathbf{x}_{n}(t)+\mathbf{v}_{n}(t)
    \end{aligned}
\end{equation}
where $\mathbf{x}_n(t)\in\mathbb{R}^{l_n}$ is the plant state vector; $\mathbf{A}_n\in\mathbb{R}^{l_n\times l_n}$ is the plant state transition matrix; $\mathbf{y}_n(t)\in\mathbb{R}^{r_n}$ is the sensor measurement vector; $\mathbf{C}_n\in\mathbb{R}^{r_n\times l_n}$ is the measurement matrix; $\mathbf{w}_n(t)\in\mathbb{R}^{l_n}$ and $\mathbf{v}_n(t)\in\mathbb{R}^{r_n}$ are the plant disturbance and sensing measurement noise vectors, respectively. These noise vectors are independent and identically distributed (i.i.d.) zero-mean Gaussian processes with covariance matrices $\mathbf{W}_n$ and $\mathbf{V}_n$, respectively.

Due to the measurement distortion and noise in \eqref{LTI}, each sensor has a local estimator to pre-estimate the plant state $\mathbf{x}_n(t)$ based on the raw measurement $\mathbf{y}_n(t)$ before sending it to the remote estimator. We adopt the classical Kalman filter (KF) for generating the estimated state $\mathbf{x}_{n}^{s}(t)$ as it is the optimal estimator of LTI systems in the average estimation MSE~\cite{Leong2020OMA}. The definition of local estimation error covariance $\mathbf{P}_n^s(t)$ is given as
\begin{equation}\label{EstimationCovariance}
    \mathbf{P}_{n}^{s}(t) \triangleq \mathbb{E}\left[\left(\mathbf{x}_{n}^{s}(t)-\mathbf{x}_{n}(t)\right)\left(\mathbf{x}_{n}^{s}(t)-\mathbf{x}_{n}(t)\right)^{\mathrm{T}}\right].
\end{equation}
Since our work focuses on the remote state estimation, each local estimator is assumed to be stable and operate in the steady state, i.e., the estimation error covariance of the local KF is a constant $\mathbf{P}_n^s(t)\triangleq{\bar{\mathbf{P}}}_n,\forall t\in\mathbb{N}$, where $\mathbb{N}$ is the set of positive integers\cite{Leong2020OMA,Liu2021FSMC,Forootani2022NOMA,liu2021deep}.

\subsection{Wireless Channel Model and Communications}\label{subsec:MarkovChannel}
We consider finite-state Markov block-fading channels \cite{Sadeghi2008FSMC}. The overall channel power gain state matrix is denoted as $\mathbf{G}(t)$, where each column vector $\mathbf{g}_{n}(t) \triangleq(g_{n, 1}(t), \ldots, g_{n, M}(t))^{\mathrm{T}}$ represents the channel power gain between sensor $n$ and the remote estimator over $M$ channels. Each channel power gain has $H$ states, i.e., $g_{n,m}(t) \in \mathcal{G} \triangleq\left\{h_{1}, h_{2}, \ldots, h_{H}\right\}$. The channel state vector $\mathbf{g}_{n}(t) \in \mathcal{G}^{M} \triangleq\left\{\tilde{\mathbf{g}}_{1}, \tilde{\mathbf{g}}_{2}, \ldots, \tilde{\mathbf{g}}_{H^{M}}\right\}$ is modeled as a multi-state Markov chain. Since sensors are dislocated and have different radio propagation environments, we assume that their Markov channel states are independent. 
The remote estimator has the knowledge of channel state information obtained via standard channel estimation techniques.
The channel state transition matrices are unknown to the remote estimator, because the estimation of a multi-dimensional Markov chain model is computationally intensive \cite{He2017FSMC}.

We adopt short-packet communications for sensor packet transmissions~\cite{Polyanskiy2010BLER}. Given the packet length $l$ (i.e., the number of symbols per packet), the number of data bits $b$, the signal-to-noise ratio (SNR) $\gamma_n$ of sensor $n$, we have the Shannon capacity $\mathcal{C}(\gamma_n)=\log_2{(1+\gamma_n)}$ and the channel dispersion $\mathcal{V}(\gamma_n)=(1-(1+\gamma_n)^{-2})(\log_2{e})^2$. Then, the decoding failure probability of sensor $n$’s packet can be approximated as \cite{Liu2021Polyanskiy}
\begin{equation}\label{BLER}
\varepsilon\left(\gamma_{n}\right) \approx \mathcal{Q}\left(\frac{\mathcal{C}\left(\gamma_{n}\right)-\frac{b}{l}}{\sqrt{\frac{\mathcal{V}\left(\gamma_{n}\right)}{l}}}\right),
\end{equation}
where $\mathcal{Q}(x)=(\frac{1}{\sqrt{2\pi}})\int_{x}^{\infty}{e^{-\frac{t^2}{2}}\text{d}t}$ is the Gaussian Q-function.

\subsection{Multiple-Access Scheme}
In the NOMA scheme, each sensor takes at most one channel for transmission, while each channel can be allocated to multiple sensors at the same time. At each time slot, the remote estimator needs to determine both the sensor-to-channel assignment and the sensor transmission power for interference management. 
Let $\mathbf{d}_n(t)\triangleq(d_{n,1}(t),d_{n,2}(t),\ldots,d_{n,M}(t))^{\mathrm{T}}\in\left\{0,1\right\}^M$ denotes the binary channel selection action of sensor $n$ at the $M$ channels, where $\sum_{m=1}^{M} d_{n, m}(t) \leq 1$.
Let $\mathbf{p}^{\mathrm{tx}}(t)\triangleq(P_1^{\mathrm{tx}}(t),P_2^{\mathrm{tx}}(t),\ldots,P_N^{\mathrm{tx}}(t))^{\mathrm{T}}\in\mathbb{P}^N$ denotes the transmission power of $N$ sensors at time $t$, where $\mathbb{P}\triangleq[0,P_{max}]$. Then, the received signal power of sensor $n$ is $P_{n}^{\mathrm{rx}}(t)=P_{n}^{\mathrm{tx}}(t)\left(\left(\mathbf{d}_{n}(t)\right)^{\mathrm{T}}\mathbf{g}_{n}(t)\right)$.

To decode sensor signals at the same channel, the remote estimator performs successive interference cancellation (SIC) with the decreasing order of the received sensor signal power~\cite{Islam2017NOMA5G}, i.e., the strongest/weakest sensor signal is decoded first/last. The first sensor packet is decoded by treating all other sensor signals as interference. Once it is decoded successfully, the sensor signal can be reconstructed perfectly and thus removed from the received signal. Then, the second sensor packet will be decoded without the interference of the first one. The decoder stops once a decoding failure occurs or the last sensor packet has been successfully decoded. 
Assuming that $P_1^{\mathrm{rx}}(t)\geq P_2^{\mathrm{rx}}(t)\geq\ldots\geq P_N^{\mathrm{rx}}(t)$, the signal-to-interference-plus-noise ratio (SINR) for decoding sensor $n$’s packet is
\begin{equation}
\gamma_{n}(t)=\frac{P_{n}^{\mathrm{rx}}(t)}{\sum_{i=n+1}^{N} \left(\mathbf{d}_{n}(t)\right)^{\mathrm{T}}\mathbf{d}_{i}(t) P_{i}^{\mathrm{rx}}(t)+\sigma^{2}},
\end{equation}
where $\sigma^2$ is the receiving noise power.
The decoding failure probability of sensor $n$ can be obtained as
\begin{equation}
\hat{\varepsilon}_{n}(t)=\!\begin{cases}
\!\varepsilon\left(\gamma_{1}(t)\right),&n=1\\
U_{n,1}+\sum_{k=2}^{n}\left(U_{n,k}\prod_{i=1}^{k-1}\left(1\!-\!U_{k,i}\right)\right),&n>1.
\end{cases}
\end{equation}
where $U_{n,n}\!=\!\varepsilon\left(\gamma_{n}(t)\right)$ and $U_{n,i}\!=\!\left(\mathbf{d}_{n}(t)\right)^{\mathrm{T}}\!\mathbf{d}_{i}(t) \hat{\varepsilon}_{i}(t),\forall i\! \leq\! n$.

\subsection{Remote State Estimation}\label{sec:sys_rem}
Due to transmission scheduling and packet detection errors, sensor $n$’s packet may not be received by the remote estimator at each time slot. Let $\zeta_n(t)\!=\!1$ denote the successful packet detection of sensor $n$ at $t$. To provide real-time state estimation of all plants, the remote estimator employs a minimum mean-square error (MMSE) state estimation for each plant \cite{Leong2020OMA,Liu2021FSMC}
\begin{equation}\label{x_hat}
\hat{\mathbf{x}}_{n}(t)=\left\{\begin{array}{cc}
\mathbf{x}_{n}^{s}(t) & \text { if } \zeta_{n}(t)=1 \\
\mathbf{A}_{n} \widehat{\mathbf{x}}_{n}(t-1) & \text { otherwise.}
\end{array}\right.
\end{equation}
Thus, the remote estimation error covariance is
\begin{align} \label{Pn_1}
\mathbf{P}_{n}(t) & \triangleq \mathbb{E}\left[\left(\hat{\mathbf{x}}_{n}(t)-\mathbf{x}_{n}(t)\right)\left(\hat{\mathbf{x}}_{n}(t)-\mathbf{x}_{n}(t)\right)^{\mathrm{T}}\right] \\
&=\left\{\begin{array}{cc}
\mathbf{\bar{P}}_{n} & \text { if } \zeta_{n}(t)=1 \\ \label{Pn_2}
\mathbf{A}_{n} \mathbf{P}_{n}(t-1) \mathbf{A}_{n}^{\mathrm{T}}+\mathbf{W}_{n} & \text { otherwise }
\end{array}\right.
\end{align}
where \eqref{Pn_2} is obtained by taking \eqref{x_hat} and \eqref{EstimationCovariance} into \eqref{Pn_1}. Recall that $\mathbf{\bar{P}}_{n}$ is the local estimation error covariance.

Now we define $\tau_n(t)$ as the age of information (AoI) of sensor $n$ at time slot $t$, representing the time interval since the last successful transmission of sensor $n$ to the remote estimator. Therefore, the AoI state of sensor $n$ has the updating rule below
\begin{equation}\label{AoI,Updating}
\tau_{n}(t)=\left\{\begin{array}{cc}
1 & \text { if } \zeta_{n}(t-1)=1 \\
\tau_{n}(t-1)+1 & \text { otherwise. }
\end{array}\right.
\end{equation}
A larger AoI indicates that the remote estimation is less accurate.
Jointly using \eqref{Pn_2} and \eqref{AoI,Updating}, the remote estimation error covariance can be concisely rewritten as a function of AoI as
\begin{equation}
\mathbf{P}_{n}(t)=f_{n}^{\tau_{n}(t)}\left(\mathbf{\bar{P}}_{n}\right),
\end{equation}
where $f_{n}^{1}(\mathbf{X})\!=\!\mathbf{A}_{n}\mathbf{X} \mathbf{A}_{n}^{\mathrm{T}}\!+\!\mathbf{W}_{n}$ and $f_{n}^{\tau_{n}}(\mathbf{X})\!=\!f_{n}^{1}\!\left(f_{n}^{\tau_{n}-1}(\mathbf{X})\right)$.

To quantify the remote estimation quality of sensor $n$ at time $t$, we define the \emph{estimation cost function}, i.e., the sum estimation MSE of the plant vector state, as
\begin{equation} \label{eq:ErrorCovariance}
\begin{aligned}
J_{n}(t)
&\triangleq 
\mathbb{E}\left[\left(\hat{\mathbf{x}}_{n}(t)-\mathbf{x}_{n}(t)\right)^{\mathrm{T}}\left(\hat{\mathbf{x}}_{n}(t)-\mathbf{x}_{n}(t)\right)\right]\\
&=\operatorname{Tr}\left(\mathbf{P}_{n}(t)\right)=\operatorname{Tr}\left(f_{n}^{\tau_{n}(t)}\left(\mathbf{\bar{P}}_{n}\right)\right),
\end{aligned}
\end{equation}
where $\operatorname{Tr}(\cdot)$ is the matrix trace operator. Thus, the AoI state and the local estimation error covariance jointly determine the remote estimation quality. A smaller $J_n(t)$ indicates that the remote state estimation is more accurate. 
As proved in our previous work~\cite{Liu2021FSMC}, if the spectral radius of plant $n$'s state transition matrix $\mathbf{A}_n$ is greater than $1$, the increasing AoI gives rise to the exponential growth of estimation cost.

\section{Problem Formulation} 
We aim to design a deterministic and stationary resource allocation policy denoted as $\pi(\cdot)\in \Pi$ that generates channel allocation and power control actions of all sensors at each time slot, for achieving the optimal discounted long-term average estimation quality \cite{Leong2020OMA}, i.e., 
\begin{equation}
J^{*}=\min _{\pi(\cdot) \in \Pi} \lim _{T \rightarrow \infty} \mathbb{E}\left[\sum_{t=0}^{T-1} \sum_{n=1}^{N} \lambda^{t} J_{n}(t)\right]
\end{equation}
where $\lambda\in(0,1)$ is a discount factor and $\mathbb{E}\left[\cdot\right]$ is the expectation operator.
We formulate such a sequential decision-making problem into an MDP. 

\subsection{MDP Formulation}\label{sec:MDP}
In general, the MDP takes both the channel quality state and the AoI state of all sensors as inputs and generates the resource allocation action at each time. The Markovian property holds directly due to the Markov channel modeling and the AoI state updating rule.

\textbf{State:} Given the channel state matrix $\mathbf{G}(t)\in\mathcal{G}^{M\times N}$, and the AoI state $\boldsymbol{\tau}(t)=(\tau_1(t),\tau_2(t),\ldots,\tau_N(t))\in\mathbb{N}^N$, the state of the MDP is defined as $\mathbf{s}\left(t\right)\triangleq\left\{\mathbf{G}(t),\boldsymbol{\tau}(t)\right\}\in\mathcal{G}^{M\times N}\times\mathbb{N}^N$.

\textbf{Action:} The hybrid action $\mathbf{a}(t)\triangleq\left\{\mathbf{D}(t),\mathbf{p}^{\mathrm{tx}}(t)\right\}$ takes into account both the discrete channel allocation $\mathbf{D}(t)$ and the continuous power control $\mathbf{p}^{\mathrm{tx}}(t)\in\mathbb{P}^N$.
The discrete action space is denoted as $\mathcal{A}$ with the cardinality of $\left|\mathcal{A}\right|=(M+1)^N$, as each sensor has $M+1$ options for channel selection. The hybrid action space is denoted as $\mathcal{A}\times\mathbb{P}^N$.

\textbf{Transition:} The state-transition probability of MDP consists of the channel state transition and the AoI state transition. Since the former does not depend on the latter, the state-transition probability from state $\mathbf{s}(t)$ to state $\mathbf{s}(t+1)$ under a particular action $\mathbf{a}(t)$ can be written as $\operatorname{Pr}[\mathbf{s}(t+1) | \mathbf{s}(t), \mathbf{a}(t)]=\operatorname{Pr}[\mathbf{G}(t+1)  |  \mathbf{G}(t)] \operatorname{Pr}[\boldsymbol{\tau}(t+1) | \mathbf{s}(t), \mathbf{a}(t)]$. Recall that the knowledge of $\operatorname{Pr}[\mathbf{G}(t+1)  |  \mathbf{G}(t)]$ is unavailable (Section~\ref{subsec:MarkovChannel}).

\textbf{Policy:} The policy is a mapping between the state and the action as $\mathbf{a}(t)=\pi(\mathbf{s}(t))$, where $\pi(\cdot)\in\Pi$.

\textbf{Reward:} We define the reward of the MDP as the negative sum estimation cost, i.e.,
$r(t)=-J(t)=-\sum_{n=1}^{N} J_{n}(t)$.
Thus, one needs to find a resource allocation policy for maximizing the discounted long-term average reward $\lim _{T \rightarrow \infty} \mathbb{E}\left[\sum_{t=0}^{T-1} \lambda^{t} r(t)\right]$.

\subsection{Challenges for Solving the MDP}\label{sec:challenge}
In the absence of state transition probabilities, conventional model-based MDP algorithms cannot work (e.g., value and policy iteration algorithms), and thus data-driven reinforcement learning approaches are preferable. Due to the curse of dimensionality introduced by the large state space  $\mathcal{G}^{N\times M}\times\mathbb{N}^N$, conventional data-driven reinforcement learning algorithms, such as Q-learning, are not applicable. 
To deal with the large state space issue, we resort to more advanced DRL algorithms adopting deep neural networks (DNNs) for value function approximation, though there are several challenges remaining.

\textbf{1) Large and hybrid action space:}
The MDP has a large action space, even with relatively small numbers of sensors and channels. For example, when $N=10, M=5$, there are $\left(M+1\right)^N=60466176$ discrete actions for channel assignment. For the continuous action part, a commonly adopted method is action discretization. In the simplest 2-level quantization scenario, there are $2^N=1024$ actions for power control and thus $1024\times60466176\approx6\times{10}^{10}$ discrete actions in total for joint channel assignment and power control.
However, deep Q-network (DQN), the most popular DRL algorithm adopted for solving MDPs with discrete actions, cannot handle such an MDP.
The large action space entails large DNNs and makes the DQN difficult to train, requiring a huge storage space and a tremendous amount of computing power. Solving MDPs with large action spaces are challenging.

\textbf{2) Training difficulties:}
In our MDP problem, the channel state is of high dynamics and the AoI state is also a stochastic function of the channel state and the action. In addition, as mentioned in Section~\ref{sec:sys_rem}, the absolute value of reward grows exponentially fast with respect to the AoI state, which means a DRL agent needs to deal with a fairly large range of rewards, inducing a highly fluctuated training process that is difficult to converge. This feature is different from many existing works applying DRL in wireless communications problems \cite{Zappone2019DRLoverview}, where the reward/cost commonly grows up in linear or log scale with the increasing state. Therefore, the highly stochastic states and the large reward range 
can lead to unstable training process and  make a DRL agent difficult to converge to an optimal policy.

\section{Deep Reinforcement Learning Algorithm} \label{sec:DRL}
\begin{figure}[t]
	\centering\includegraphics[width=3.4in]{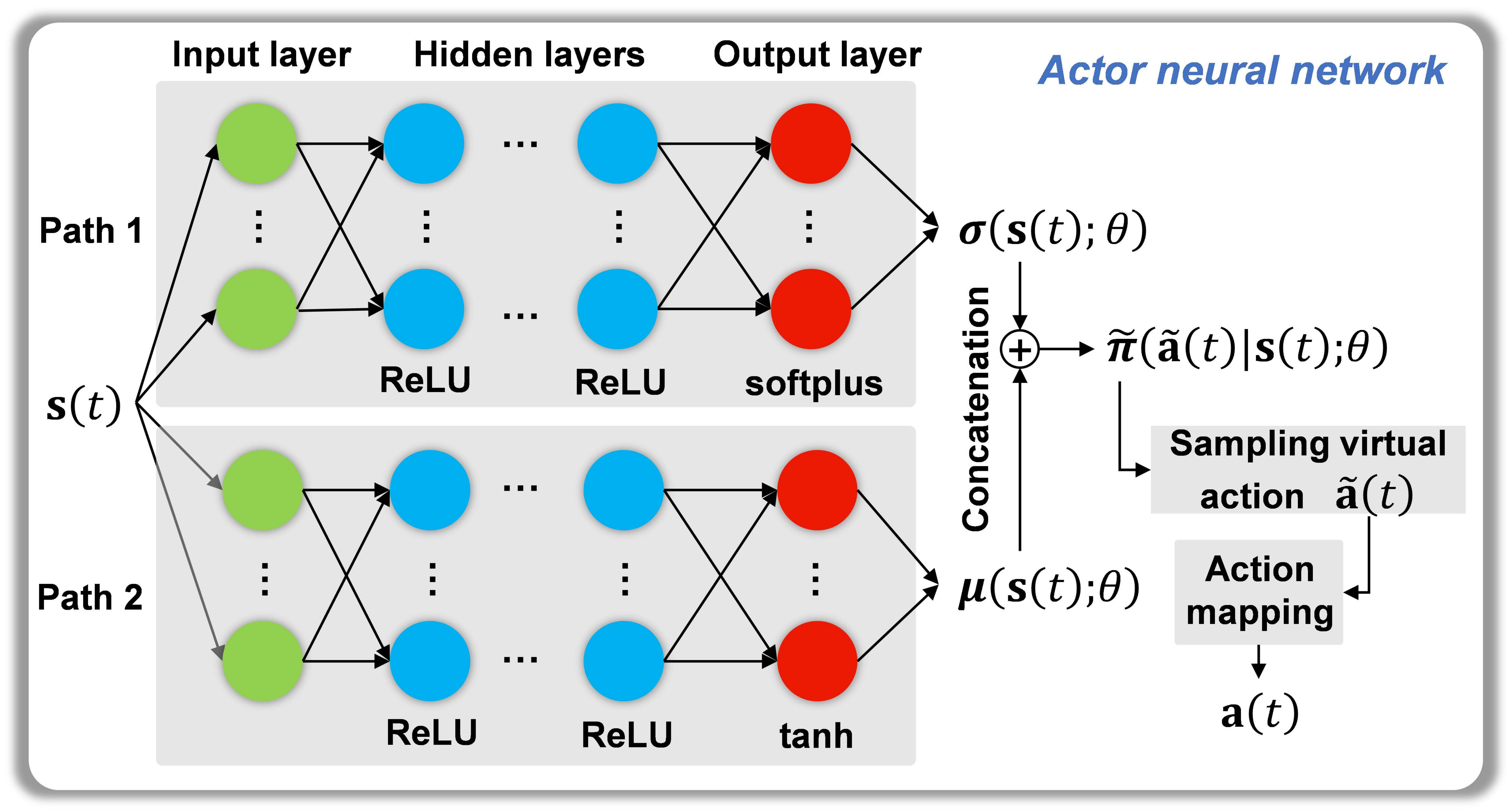}
	\vspace{-0.3cm}	
	\caption{The framework for generating resource allocation actions based on a trained actor NN (Section~\ref{subsec:PPO}) and the action mapping scheme (Section~\ref{sec:ActionMapping}).}
	\label{fig:ActionMapping}
	\vspace{-1.5em} 
\end{figure}

To solve the challenges above, first, we design low-dimensional continuous virtual actions of the resource allocation problem and propose a novel scheme that maps virtual action $\tilde{\mathbf{a}}(t)$ to real hybrid discrete and continuous action $\mathbf{a}(t)$. Second, we choose a proper DRL framework with a continuous action space, which provides a stable training process to learn the optimal policy with virtual actions. The framework is illustrated in Fig.~\ref{fig:ActionMapping}.

\subsection{Virtual Action Design and Action Mapping Scheme} \label{sec:ActionMapping}
Considering the hybrid action feature of resource allocation, it is natural to design a multi-dimensional continuous virtual action of each sensor to represent the hybrid action. However, how to design a continuous virtual action and map it to a discrete channel selection action for effective DRL is highly non-trivial. A naive method is to adopt a scalar continuous virtual action for each sensor and discretize it linearly into $M+1$ levels for channel selection. Level 0 means no channel selection and Level $m>0$ denotes the selection of channel $m$. At first glance, the method compresses the $M$-size discrete action space by a one-dimensional continuous one. However, the method implicitly converts the original non-Euclidean action space into the Euclidean one. This potentially requires the DRL agent to learn a highly discontinuous multi-level-multi-stage piecewise policy function. Ideally, a slight change of state can make a significant change in action output. This makes the DRL difficult to converge to an optimal one.

To solve the issue of discontinuous function approximation, we propose to convert the decimal channel selection action (i.e., from $0$ to $M$) of each sensor to a binary sequence with a length of $\lceil \log_{2}(M+1) \rceil$. Thus, we design the virtual action for each sensor with $\lceil \log_{2}(M+1) \rceil $ elements. Each element is a real number, and the positive and negative values are mapped to ‘1’ and ‘0’, respectively, for real action mapping. Now, the virtual action has $N \lceil \log_{2}(M+1) \rceil$ dimension for channel selection of all sensors. Each element of the virtual action only handles the selection of two discrete actions. Changing of a single virtual action entry from negative to positive represents the reduced and increased likelihood of one action and the other, respectively. The continuity can make the DRL easy to train. The original $(M+1)^N$-size discrete action space is compressed by the $N \lceil \log_{2}(M+1) \rceil $-dimension continuous one, which scales linearly and logarithmically in $N$ and $M$, respectively.
One dimensional virtual action $\tilde{P}^{\mathrm{tx}}_n(t) \in \mathbb{R}$ is added for each sensor’s transmit power control, and the mapping between the virtual to real power control action is
$P^{\mathrm{tx}}_n(t) =P_{\max} (\operatorname{clamp_{-1}^1}[\tilde{P}^{\mathrm{tx}}_n(t)]+1)/2$, where the operator $\operatorname{clamp_{-1}^1}[\cdot]$ is for truncating a variable to values between $-1$ and $1$.
Thus, the virtual action $\tilde{\mathbf{a}}(t)$ has $N \lceil \log_{2}(M+1) \rceil +1$ elements in total.

\subsection{Policy Optimization with Virtual Actions}\label{subsec:PPO}
There are many DRL frameworks with various features developed during the past decade. 
To match the key problem features in Section~\ref{sec:challenge}, we adopt the proximal policy optimization (PPO) \cite{Schulman2017PPO}, a \emph{policy-based} DRL method that learns a \emph{stochastic policy} via \emph{on-policy learning}, for solving the MDP. The reasons include: 1) compared with off-policy DRL methods, on-policy ones are more stable in highly stochastic environments due to the guaranteed monotonic performance improvement based their policy updating mechanism \cite{Yang2020DeterMDP}; 2) existing research shows that stochastic policy based DRL algorithms lead to smoother function approximations \cite{Moarales2020Book}, and thus provide higher speed for training convergence and better training stability than deterministic ones when the training environment is of high dynamics \cite{Xiao2021DRLpolicy}; 3) policy-based DRL generates multi-dimensional continuous actions directly and can deal with MDPs with continuous action spaces.

The PPO agent has a pair of actor and critic neural networks (NNs), where the NN parameters, including weights and biases, are denoted by $\theta$ and $\varphi$, respectively. Each of the NNs has an input layer, multiple hidden layers, an output layer, and adopts a fully connected and feedforward structure. The input layers of the actor and critic NNs receive the MDP state $\mathbf{s}(t)$. All hidden layers adopt the activation function of rectified linear unit (ReLU) for nonlinear function approximation.

We design the actor NN for generating a multi-dimensional \textit{continuous virtual action} $\tilde{\mathbf{a}}(t)$ with stochastic policy $\tilde{\pi}(\tilde{\mathbf{a}}(t)|\mathbf{s}(t);\theta)$, which is a probability density function of $\tilde{\mathbf{a}}(t)$ given the current state $\mathbf{s}(t)$. The virtual action is then mapped to the real action $\mathbf{a}(t)$ for radio resource allocation as discussed in Section~\ref{sec:ActionMapping} to obtain the next state $\mathbf{s}(t+1)$ and reward $r(t+1)$. The actor output layer generates mean $\boldsymbol{\mu}(\mathbf{s}(t);\theta)$ and standard deviation $\boldsymbol{\sigma}(\mathbf{s}(t);\theta)$ of $\tilde{\mathbf{a}}(t)$. We adopt tanh and softplus as the activation functions for the mean and the standard deviation outputs, respectively. The former is to bound the mean within $[-1,1]$, while the latter is to guarantee a positive standard deviation. The critic NN estimates the state-value function $V(\mathbf{s}(t);\varphi)$ given the actor’s policy $\tilde{\pi}(\tilde{\mathbf{a}}(t)|\mathbf{s}(t);\theta)$, i.e.,
\begin{equation}
V(\mathbf{s}(t) ; \!\varphi)\!\approx\!\mathbb{E}\!\left[\sum_{k=0}^{\infty} \lambda^{k} r(t\!+k) \bigg| \tilde{\pi}(\tilde{\mathbf{a}}(t\!+k) | \mathbf{s}(t\!+k) ;\! \theta),\! \forall k\! \geq\! 0\right]\!\!.
\end{equation}
Thus, the critic NN has a single output.

The training of the PPO agent alternates between experience generation and policy update. In the following, we only present the key steps, and the detailed algorithm can be found in~\cite{Schulman2017PPO}.

1) Experience generation. By using current policy $\tilde{\pi}(\cdot|\cdot;\theta_{\text{old}})$, the PPO agent samples data $(\mathbf{s}(t),\tilde{\mathbf{a}}(t),r(t))$ with length of $T$ through interacting with the environment.
By leveraging the generated experience, the advantage function $A(t)$ and the reward-to-go function $R(t)$ for each $t=0,\dots,T-1$ can be calculated as
\begin{equation}\label{AdvantageFunc}
A(t)=\sum_{k=t}^{T-1}(\lambda \alpha)^{k-t} (r(k)+\lambda V(\mathbf{s}(k+1) ; \varphi)-V(\mathbf{s}(k) ; \varphi))
\end{equation}
and 
\begin{equation}\label{reward-to-go}
\begin{aligned}
R(t)= r(t) + \lambda V(\mathbf{s}(t+1) ; \varphi)
\end{aligned}
\end{equation}
respectively, where  $\alpha$ is a hyper-parameter named as the smoothing factor.

2) Policy update.
The PPO agent creates a mini-batch data set by randomly sampling $B$ data from the generated $T$-length experience earlier, i.e., $\{(\mathbf{s}(t_i),\tilde{\mathbf{a}}(t_i), A(t_i), R(t_i))\}$ where $t_i \in \{t_1,\dots,t_B\} \subset \{0,\dots,T-1\}$.

The loss function for updating the critic NN is defined as
\begin{equation}\label{eq:LossForCritic}
L_{\text {C}}(\varphi)=\frac{1}{B} \sum_{i=1}^{B}\left(R(t_i)-V\left(\mathbf{s}(t_i) ; \varphi\right)\right)^{2},
\end{equation}
which is named as the temporal difference error, specifying difference of the state-value estimations based on time steps $t_i$ and $t_i+1$.
The loss function for updating the actor NN has been elaborately designed for achieving high training stability and is much more complex:
\begin{equation}\label{eq:LossForActor}
\begin{aligned}
&L_{\text {A}}(\theta)\!=\\
&\frac{1}{B}\!\! \sum_{i=1}^{B}\!\left(\min\!\left\{p(\mathbf{s}(t_i);\theta)\!A(t_i), c\left(\mathbf{s}(t_i) ; \theta\right)\!A(t_i)\right\}\!+\!w \hbar_{\theta_{\text {old }}}\!\!\left(\!\tilde{\mathbf{a}}(t_i) \right)\right)
\end{aligned}
\end{equation}
where $p(\mathbf{s}(t_i);\theta) = \frac{\tilde{\pi}(\tilde{\mathbf{a}}(t_i)|\mathbf{s}(t_i) ; \theta)}{\tilde{\pi}\left(\tilde{\mathbf{a}}(t_i)| \mathbf{s}(t_i) ; \theta_{\text {old }}\right)}$ is a density ratio, and 
$
c(\mathbf{s}(t_i);\theta)=\max \left\{\min \left\{p(\mathbf{s}(t_i);\theta), 1+\omega\right\}, 1-\omega\right\}
$ is a clip function with a hyper-parameter $\omega$. 
$\hbar_{\theta_{\text {old}}}\!\!\left(\tilde{\mathbf{a}}(t_i) \right)$ is the entropy loss function of $\tilde{\mathbf{a}}(t_i)$. Since $\tilde{\mathbf{a}}(t_i)$ follows a Gaussian distribution, its entropy loss can be directly obtained based on its standard deviation $\boldsymbol{\sigma}(\mathbf{s}(t_i);\theta_{\text{old}})$.
$w$ denotes entropy loss weight factor. 
Then, the critic NN and the actor NN parameters $\varphi$ and $\theta$ can be updated by optimizing the loss functions $L_{\text {C}}(\varphi)$ and $L_{\text {A}}(\theta)$, respectively, using the widely adopted Adam optimizer.

After training, the virtual action $\tilde{\mathbf{a}}(t)$ can be generated deterministically based on the maximum likelihood method for online deployment, i.e., $\tilde{\mathbf{a}}(t) = \boldsymbol{\mu}(\mathbf{s}(t);\theta)$.

\section{Numerical Experiments} \label{sec:simulation}
\subsection{Experiment Setup}
Our numerical experiments are implemented on a computing platform with two Intel Xeon Gold 6256 CPUs @ 3.60 GHz and a 192 GB RAM. Each of the actor and critic NNs of the PPO-based DRL agent has three hidden layers with sizes of $\left\lceil70K\right\rceil, \left\lceil50K\right\rceil,\left\lceil30K\right\rceil$, respectively, where $K=\sqrt{N/M}\log _{2}(M+1)$. The state input of each NN has $N(M+1)$ dimensions, which is the same as the MDP. The output size of the critic NN is $1$, while that of the actor NN is discussed in Section~\ref{subsec:PPO}. The dynamic system matrices $\mathbf{A}_n$ are randomly generated by leveraging the method presented in \cite{Leong2020OMA}, where the spectrum radius is drawn uniformly from the range of $(1, 1.3)$. The channel transition matrices are generated randomly. Table~\ref{tab:Setup} summarizes the details of the remote estimation system parameters and the DRL parameters.
\begin{table}[t]
\footnotesize
\setlength\tabcolsep{7.5pt}
\centering
\caption{Summary of Experiment Setup}
	\vspace{-0.3cm}
\begin{tabular}{cc}
\hline\hline
Items   & Value    \\ \hline
\multicolumn{2}{l}{\textit{\textbf{Remote state estimation system parameters}}}\\
\rowcolor[HTML]{EFEFEF} 
Transmit power budget {[}dBm{]}, $P_{max}$  & 23 \\
Receiving noise power {[}dBm{]}, $\sigma^2$    & $-$60   \\
\rowcolor[HTML]{EFEFEF} 
Code rate {[}bps{]}, $b/l$    & 2        \\
Block length {[}symbols{]}, $l$   & 200    \\
\rowcolor[HTML]{EFEFEF} 
Markov channel power gain states, $\mathcal{G}$  &  $\{10^{-8},10^{-7},\cdots,10^{-1}\}$ \\
Channel state transition matrix  & Randomly generalized  \\ \hline
\multicolumn{2}{l}{\textit{\textbf{Traning parameters}}} \\ 
\rowcolor[HTML]{EFEFEF} 
Episode number, $E$    &  $\left\lceil 250 \times \frac{N}{M} \times \sqrt{NM} \right\rceil$ \\
Maximum time steps per episode, $T$     & 128       \\
\rowcolor[HTML]{EFEFEF} 
Learning rate of actor NN    & 0.0001  \\
Learning rate of critic NN   & 0.001     \\
\rowcolor[HTML]{EFEFEF} 
Mini-batch size, $B$    & 128      \\
Discount factor, $\lambda$     & 0.95    \\\hline
\multicolumn{2}{l}{\textit{\textbf{Learning agent parameters of this work (PPO)}}} \\
\rowcolor[HTML]{EFEFEF} 
The smoothing factor, $\alpha$ & 0.95  \\
Entropy loss weight factor, $w$ & 0.01  \\
\rowcolor[HTML]{EFEFEF} 
Clip factor, $\omega$   & 0.2    \\ \hline
\multicolumn{2}{l}{\textit{\textbf{Learning agent parameters of the benchmark (DQN)}}}\\
\rowcolor[HTML]{EFEFEF}
Initial epsilon for exploring action space    & 1\\
Epsilon decay rate    & 0.999 \\
\rowcolor[HTML]{EFEFEF}
Minimum epsilon  & 0.01 \\
Experience buffer length   & $1000NM$    \\\hline\hline
\end{tabular}
\label{tab:Setup}
\vspace{-1.5em} 
\end{table}

We adopt the DQN-based channel assignment of a remote estimation system with OMA as a benchmark. 
As shown in~\cite{Leong2020OMA}, the DQN-based algorithm performs better than 
the heuristic algorithms (e.g., the greedy and the round-robin policies). Thus, we only need to compare our algorithm with the DQN-based one. 
In addition, we use the naive action mapping based PPO (discussed in Section~\ref{sec:ActionMapping}) as a benchmark of our proposed novel action mapping scheme.

\subsection{Performance Evaluation}
In Table~\ref{tab:scalability}, we compare the remote estimation performance, i.e., the average sum MSE of all plants,  between the proposed DRL-based algorithm and the benchmarks with various system scales and sensor-to-channel ratios (SCRs).
The remote estimation system with a smaller average estimation MSE has a better  performance.
Average estimation MSEs in Table~\ref{tab:scalability} are calculated by $10000$-step simulations.
\begin{table}[t]
\footnotesize
\setlength\tabcolsep{9pt}
\centering
\caption{Performance Comparison of the Proposed Algorithm and the Benchmarks in Terms of Averaged Estimation MSE}
\vspace{-0.3cm}
\begin{tabular}{c|c|cc}
\hline\hline
System scale & OMA & \multicolumn{2}{c}{NOMA} \\ \hline
($N,M$, SCR) & DQN & Naive action mapping & This work \\ \hline
\rowcolor[HTML]{EFEFEF} 
(6, 3, 2) & 46.6243 & 39.0462 & 38.0663  \\
(10, 5, 2) & $-$ & 65.0766 & 63.0768 \\
\rowcolor[HTML]{EFEFEF} 
(20, 10, 2) & $-$ & 154.9627 & 125.4118 \\
(30, 15, 2) & $-$ & 297.5236 & 218.3914 \\
\rowcolor[HTML]{EFEFEF} 
(40, 20, 2) & $-$ & 397.5561 & 286.1228\\
(50, 25, 2) & $-$ & 486.9513 & 360.9214 \\ \hline
\rowcolor[HTML]{EFEFEF} 
(10, 4, 2.5) & $-$ & 78.8600 & 77.6075 \\
(20, 8, 2.5) & $-$ & 172.9613 & 157.2169  \\
\rowcolor[HTML]{EFEFEF} 
(30, 12, 2.5) & $-$ & 318.8591 & 252.8929  \\
(40, 16, 2.5) & $-$ & 404.8909 & 345.0570 \\
\rowcolor[HTML]{EFEFEF} 
(50, 20, 2.5) & $-$ & 563.4724 & 434.4692 \\ \hline
(15, 5, 3) & $-$ & 159.5629 & 129.8431 \\
\rowcolor[HTML]{EFEFEF} 
(24, 8, 3) & $-$ & 275.5904 & 230.7575 \\
(33, 11, 3) & $-$ & 438.7793 & 343.1833 \\
\rowcolor[HTML]{EFEFEF} 
(42, 14, 3) & $-$ & 578.4324 & 448.6448 \\
(51, 17, 3) & $-$ & 765.4839 & 592.1996\\ \hline\hline
\end{tabular}
\label{tab:scalability}
\vspace{-1.5em} 
\end{table}

We see that the DQN algorithm with OMA only works for the $6$-sensor-$3$-channel setting and does not even converge for larger systems. The proposed DRL algorithm with NOMA can scale up to $50$ sensors and $25$ channels.
We also see that the proposed action mapping scheme-based algorithm can provide a $25$\% average estimation MSE reduction than the naive mapping scheme-based one when the system is large (e.g., $N\geq 50$).
The performance gap increase with the growing system scale.
We also see that the estimation performance decreases with the increasing SCR as expected, due to the decreasing amount of wireless resource.

\section{Conclusion}\label{sec:conclusion}
We have proposed a practical remote estimation system with the NOMA scheme. We have developed an advanced DRL algorithm for resource allocation with large hybrid state and action spaces. Our experiments have showcased that the proposed DRL algorithm is able to effectively address the resource allocation problem and provide significant performance gain than the benchmarks.
For future work, we will investigate distributed DRL algorithms for resource allocation of large-scale systems and compare them with the present centralized allocation scheme.


\end{document}